\documentclass[reprint,aps,pra,longbibliography]{revtex4-2}

\usepackage[T1]{fontenc}
\usepackage[utf8]{inputenc}
\usepackage{graphicx}
\usepackage{amsmath,amssymb}
\usepackage{siunitx}
\usepackage{hyperref}
\usepackage{orcidlink}
\usepackage{xcolor}
\colorlet{green}{green!60!black}
\usepackage{tikz}
\usetikzlibrary{calc,arrows.meta,decorations.pathmorphing,positioning}

\hypersetup{
  colorlinks=true,
  linkcolor=blue,
  citecolor=blue,
  urlcolor=blue
}

\usepackage{pgfplots}
\pgfplotsset{compat=1.18}

\begin{document}

\title{Traction Constraints and the Physics of Faster-Than-the-Wind Travel}

\author{Karl Svozil \orcidlink{0000-0001-6554-2802}}
\affiliation{Institute for Theoretical Physics, TU Wien,
Wiedner Hauptstra{\ss}e 8-10/136, 1040 Vienna, Austria}
\email{karl.svozil@tuwien.ac.at}

\begin{abstract}
It is a well-documented yet counterintuitive fact that wind-driven vehicles (with no onboard power source) can travel directly downwind faster than the wind itself. This effect is not paradoxical once one recognizes that the vehicle is not pushed by the air alone but acts as a coupled mechanical system that taps the relative motion of two media---moving air and stationary ground (or, for watercraft, water taken as quiescent in the far field, neglecting currents)---and, through its drivetrain, can transform a modest velocity difference into a larger vehicle speed. The essential ingredient is a rigid constraint: the wheel--ground contact enforces a no-slip rolling (traction) constraint and supplies tangential reaction forces. In the ideal limit this contact does no work in the ground frame because the instantaneous contact-point velocity is zero; dissipation enters only through aerodynamic drag, rolling resistance, bearing losses, and slip. The drivetrain (wheels, gears, propeller) then acts as a mechanical transformer, trading force against speed in the usual way so that power is conserved in the lossless limit. Using the analogies of a gearbox, a lever, and a sliding-boat thought experiment, this work gives an explicitly Newtonian description of how faster-than-the-wind travel arises from coupling two media through traction constraints and a transmission.
\end{abstract}

\maketitle

\section{Introduction: A Counter-Intuitive Phenomenon}

The ability of a wind-powered vehicle to travel directly downwind faster than the wind itself appears paradoxical~\cite{Blackford1978,Stong1975,McDonald2021}. At first glance, such motion seems to violate energy conservation---like pulling oneself up by one's own bootstraps. Yet this is not a theoretical curiosity but an experimentally verified reality. The principle was analyzed as early as 1969 by Bauer~\cite{Bauer1969,bauer1971sailing,bauer1974taking}, and later by  Gaunaa et al.~\cite{Gaunaa2009} and  McDonald~\cite{McDonald2021} among others~\cite{khan2013analysis,AAPT2013}.

The ``Blackbird'' land yacht~\cite{nahon-2023}, for instance, has been clocked at 2.8 times the true wind speed downwind~\cite{nalsa2010}, and, with reverse gearing, exceeded twice the wind speed upwind. The phenomenon is robust enough to have served as a problem for the United States Physics Olympiad~\cite{AAPT2013}. Despite this established literature, the counter-intuitive nature of the effect persists, recently sparking a high-profile public wager among physicists~\cite{Veritasium2021}.

In Section~\ref{sec:pedagogy} we make the educational value explicit and provide a practical guide for classroom and laboratory use.
The misunderstanding arises from picturing the vehicle as a passive body pushed by the air alone. If that were the case, its speed could never exceed that of the wind \emph{when traveling directly downwind}.

A conventional sailboat can exceed the true wind speed on a reach, and can exceed the wind in downwind velocity-made-good by sailing at an angle and jibing. However, sustaining a speed exceeding the true wind \emph{directly dead downwind} requires a mechanism that can generate thrust even when the apparent wind vanishes or reverses (e.g., a drivetrain-coupled propeller), which is the regime addressed here.

The crucial ingredient is the ground (or water) contact. In what follows we treat it explicitly as a \emph{constraint}: the wheel--ground interface enforces rolling without slip and can therefore transmit tangential reaction forces (traction). In the ideal rolling limit the ground does not supply mechanical power in the ground frame because the instantaneous velocity of the contact point is zero; instead, the contact provides the reaction forces and torques needed to transmit power internally through the drivetrain. Non-ideal effects enter through aerodynamic drag, rolling resistance, bearing losses, and any slip at the contact.

A recurring theme throughout this paper is that sustained faster-than-the-wind
motion requires not merely a moving airstream, but a coupling between two media
in relative motion together with an external reaction constraint. On land this
constraint is provided by the traction-enabled wheel--ground contact; on water
it is provided more weakly by hydrodynamic reaction, with unavoidable slip and
wake losses. Later sections return to this point in specific contexts, but the
basic mechanism is always the same: the vehicle acts as a mechanical
transformer that exchanges momentum between the two media through an external
constraint.

We first describe how the vehicle operates in both downwind and upwind configurations and clarify why the same mechanism does not apply to aircraft, which interact with only a single medium. We then interpret the physics through three complementary mechanical analogies---a gearbox, a lever, and a ``sliding boat'' thought experiment---that make explicit the role of external constraints. In particular, sustained velocity amplification and thrust require an external reaction surface (in practice, a traction-enabled, approximately no-slip wheel--ground contact) that allows the drivetrain to exchange momentum between air and ground.

\section{The Physics of Faster-Than-the-Wind Travel}

Throughout, ``faster-than-the-wind'' refers to motion parallel (upwind or downwind) to the true wind in the ground/water (laboratory) frame \(G\), unless stated otherwise.
In this paper ``wind-driven'' means that no auxiliary onboard energy source
(engine, battery, spring, or flywheel) supplies net work; apart from start-up
transients, sustained motion is powered solely by the wind (air motion relative
to the ground/water).

We use a single velocity nomenclature throughout the paper. The true wind, or
the moving upper rack in the gearbox analogy, has speed \(v_w\) in the
laboratory frame \(G\).
The vehicle, boat, or gearbox casing has signed velocity component \(v\)
along the wind axis in \(G\). Positive \(v\) denotes motion with the wind,
and negative \(v\) denotes motion against the wind.
The apparent-wind velocity in the vehicle frame \(C\) is always denoted
by \(v_a=v_w-v\). Thus the same symbols are used in the cart, watercraft,
gearbox, lever, and pedagogical discussions; only the physical realization of
the constraint changes.

\paragraph*{Reference frames and sign conventions.}
To avoid ambiguity, we use only two frames unless explicitly stated otherwise.
The first is the ground/water (laboratory) frame \(G\), in which the ground is
at rest and, for boats, the far-field water is at rest (neglecting currents).
All quoted vehicle speeds and all statements such as ``faster than the wind''
refer to \(G\). The second is the instantaneous vehicle frame \(C\). Because
the vehicle may accelerate in \(G\), \(C\) is understood as the momentarily
comoving frame at a given instant; we use it only for local kinematics and sign
conventions, not as a global frame for the entire motion. For motion along the
wind direction, the apparent-wind velocity in \(C\)---more precisely, the
component of the relative air velocity along the direction of travel---is
\[
v_a = v_w - v,
\]
where \(v_w\) is the true wind speed in \(G\) and \(v\) is the vehicle speed.
Thus \(v_a>0\) means the air approaches from behind, \(v_a=0\) at the crossover
\(v=v_w\), and \(v_a<0\) means the air approaches from ahead in the vehicle
frame. We use \(G\) for the overall kinematics, momentum and power balances,
and the no-slip condition at the wheel--ground contact. We use \(C\) only to
describe the local operating state of the propeller and, in the gearbox
analogy, to fix the sign of tangential velocities before returning to \(G\). In
particular, statements such as ``the ground does no work at the contact'' are
statements in \(G\), since contact power is frame-dependent.

At the heart of faster-than-the-wind motion lies a transmission that couples two distinct media---the air and the ground.
Through its propeller and wheels, the vehicle exerts forces on both the air
and the ground, transferring momentum between them via the drivetrain. The relative motion between air and ground provides the energetic reservoir that the vehicle can tap. The instantaneous net mechanical power transferred (and thus available for acceleration) depends on both this relative velocity and the coupling forces set by propeller/wheel loading, gearing, drag, and losses.
As will be discussed later, this is analogous to a gear train or lever: It gears up velocities and gears down forces, so that the product of force and speed (power) is preserved in the ideal limit.

The direction of net mechanical power transfer through the drivetrain---that is, whether the propeller delivers shaft power to the wheels (turbine mode) or the wheels deliver shaft power to the propeller (fan mode)---is set by the gearing and blade pitch (for a given operating point).
Locally, power is distributed among components and losses, but the sign of the net shaft power between propeller and wheels distinguishes the two regimes.

In the usual downwind-faster-than-the-wind configuration, let the cart speed in
the ground frame \(G\) be \(v\), with true wind speed \(v_w\). The propeller
then experiences the apparent-wind velocity \(v_a=v_w-v\) in the vehicle frame
\(C\). Once rolling, the wheels drive the propeller, which acts as a fan and
accelerates air rearward. At the crossover \(v=v_w\), the apparent-wind
velocity vanishes (\(v_a=0\)); for \(v>v_w\), \(v_a<0\),
meaning that in the vehicle frame \(C\) the air now arrives from ahead.
Nevertheless, the propeller
continues to operate as a
wheel-driven fan and accelerates this air rearward,
thereby producing thrust. The sign change of the apparent-wind velocity at the
crossover is illustrated in Fig.~\ref{fig:dimensionless_plots}(b).
A fan consumes mechanical power: In this downwind case the propeller operates in
\emph{fan mode} and therefore requires positive shaft power. That shaft power is
delivered through the drivetrain from the rolling wheels, while the \emph{ultimate}
energy source remains the wind (air motion relative to the ground). The fan
imparts rearward momentum to the air and produces a pressure rise across the
propeller disk, thereby generating thrust.

In the upwind configuration (obtained by reversing the gearing or blade pitch),
the wind drives the propeller as a turbine, which in turn drives the wheels
against the ground. A turbine extracts mechanical power: It converts the kinetic
energy of the moving air (the wind) into shaft power, which is transmitted
through the drivetrain to the wheels and propels the vehicle upwind.

In both configurations the vehicle as a whole is ultimately powered by the wind:
It taps the kinetic energy of the air relative to the ground. However, in the
downwind-faster-than-the-wind mode the propeller itself operates as a
wheel-driven fan (shaft power delivered to the air), whereas in the upwind mode
it operates as a turbine (shaft power extracted from the air).

\paragraph*{Relation to Bernoulli's principle.}
The present explanation is intentionally Newtonian: the essential mechanism is
momentum and power transfer between two media through the drivetrain and the
traction constraint at the ground contact. For readers who prefer a fluid
description, Bernoulli's principle provides a complementary account of the
pressure field associated with the accelerated flow. In an ideal actuator-disk
picture of the propeller, Bernoulli's equation may be applied separately along
streamlines upstream and downstream of the disk, while across the disk itself
there is a pressure jump and a change in mechanical energy: in fan mode the
disk adds energy to the air, whereas in turbine mode it extracts energy from
it. Thus Bernoulli helps describe the accompanying pressure differences, but it
does not by itself explain faster-than-the-wind travel; the essential point
remains the mechanical coupling of air and ground (or water) through an
external reaction constraint.

\subsection{Downwind Faster Than the Wind}

\begin{figure*}[htbp]
\centering
\begin{tabular}{ccc}
\begin{tikzpicture}[
    scale=0.85, transform shape,
    line cap=round,
    line join=round,
    >=Stealth,
    font=\normalsize
  ]
  \useasboundingbox (-2.2,-0.7) rectangle (8.0,5.7);

  \def\rW{1.02}
  \coordinate (W) at (0,\rW);     
  \coordinate (P) at (5.25,4.00); 

  \def\thS{28.5}                   
  \pgfmathsetmacro{\thB}{\thS+180} 

  \draw[thick] (-2.0,0) -- (2.3,0);
  \node[below] at (2.25,0) {ground};
  \foreach \x in {-1.75,-1.05,-0.35,0.35,1.05,1.75}{
    \draw (\x,0) -- ++(-0.45,-0.38);
  }

  \foreach \y in {4.80,4.05,3.30}{
    \draw[->,thick,decorate,decoration={snake,amplitude=2pt,segment length=18pt}]
      (-0.85,\y) -- (1.30,\y);
  }
  \node[anchor=west] at (1.35,4.05) {wind};

  \draw[thick] (W) circle (\rW);
  \draw[thick] ($(W)+(-0.18,-0.18)$) rectangle ($(W)+(0.18,0.18)$);
  \node at ($(W)+(0,-0.45)$) {gear};

  \draw[thick] (W) -- (P);

  \draw[->,thick] ($(W)+(205:1.45)$)
    arc[start angle=205,end angle=125,radius=1.45];

  \begin{scope}[shift={(P)}, rotate=\thB]
    \path[draw=black,thick,fill=white]
      (0,0)
      .. controls (0.10,0.10) and (0.50,0.65) .. (0.32,1.28)
      .. controls (0.02,1.58) and (-0.28,0.92) .. (-0.10,0.12)
      -- cycle;
  \end{scope}

  \begin{scope}[shift={(P)}, rotate={\thB+180}]
    \path[draw=black,thick,fill=gray!45]
      (0,0)
      .. controls (0.10,0.10) and (0.50,0.65) .. (0.32,1.28)
      .. controls (0.02,1.58) and (-0.28,0.92) .. (-0.10,0.12)
      -- cycle;
  \end{scope}

  \fill (P) circle (1.4pt);

  \draw[->,thick] ($(P)+(65:0.95)$)
    arc[start angle=65,end angle=-10,radius=0.95];

  \node at ($(P)+(0.55,-1.8)$) {propeller};

\end{tikzpicture}
&&
\begin{tikzpicture}[
    scale=0.85, transform shape,
    line cap=round,
    line join=round,
    >=Stealth,
    font=\normalsize
  ]
  \useasboundingbox (-2.2,-0.7) rectangle (8.0,5.7);

  \def\rW{1.02}
  \coordinate (W) at (0,\rW);     
  \coordinate (P) at (5.25,4.00); 

  \def\thS{28.5}                   
  \pgfmathsetmacro{\thB}{\thS+180} 

  \draw[thick] (-2.0,0) -- (1.65,0);
  \node[below] at (2.15,0) {ground};
  \foreach \x in {-1.75,-1.05,-0.35,0.35,1.05,1.55}{
    \draw (\x,0) -- ++(-0.45,-0.38);
  }

  \foreach \y in {4.80,4.05,3.30}{
    \draw[->,thick,decorate,decoration={snake,amplitude=2pt,segment length=18pt}]
      (-0.85,\y) -- (1.30,\y);
  }
  \node[anchor=west] at (1.35,4.05) {wind};

  \draw[thick] (W) circle (\rW);
  \draw[thick] ($(W)+(-0.28,-0.28)$) rectangle ($(W)+(0.28,0.28)$);
  \draw[thick] ($(W)+(-0.16,-0.16)$) rectangle ($(W)+(0.16,0.16)$);
  \node[align=center] at ($(W)+(0,+0.52)$) {reverse};
  \node[align=center] at ($(W)+(0,-0.62)$) {gear};

  \draw[thick] (W) -- (P);

  \draw[->,thick] ($(W)+(125:1.45)$)
    arc[start angle=125,end angle=205,radius=1.45];

  \begin{scope}[shift={(P)}, rotate=\thB]
    \path[draw=black,thick,fill=white]
      (0,0)
      .. controls (0.10,0.10) and (0.50,0.65) .. (0.32,1.28)
      .. controls (0.02,1.58) and (-0.28,0.92) .. (-0.10,0.12)
      -- cycle;
  \end{scope}

  \begin{scope}[shift={(P)}, rotate={\thB+180}]
    \path[draw=black,thick,fill=gray!45]
      (0,0)
      .. controls (0.10,0.10) and (0.50,0.65) .. (0.32,1.28)
      .. controls (0.02,1.58) and (-0.28,0.92) .. (-0.10,0.12)
      -- cycle;
  \end{scope}

  \fill (P) circle (1.4pt);

  \draw[->,thick] ($(P)+(65:0.95)$)
    arc[start angle=65,end angle=-10,radius=0.95];

  \node at ($(P)+(0.55,-1.8)$) {propeller};

\end{tikzpicture}
\\
(a) Downwind Operation && (b) Upwind Operation
\end{tabular}
\caption{Schematic of a wind-driven vehicle's operation.  Arrows indicate the direction of rotation. The pitch of the propeller blades is indicated by the darker shade of the blade facing the rear.
(a) For downwind travel faster than the wind, the wheels drive the propeller, which pushes air backward to generate thrust.
(b) For upwind travel, the wind turns the propeller (acting as a turbine), which drives the wheels and pulls the vehicle against the wind.}
\label{fig:vehicle_operation}
\end{figure*}

Consider a cart with a propeller linked to its wheels by a chain and gears, as shown in Fig.~\ref{fig:vehicle_operation}(a). When stationary, a tailwind exerts a force on the cart and on the propeller blades, initiating motion. As the wheels turn, they drive the propeller so that it actively pushes air \emph{backward} relative to the cart. The propeller thus acts not as a passive windmill but as a fan powered by the rolling motion of the wheels against the ground.

As the cart accelerates and its ground speed approaches the wind speed, the apparent-wind velocity at the propeller diminishes. Even when the apparent-wind velocity is momentarily zero, the rolling wheels can still deliver shaft power to the propeller, which continues to accelerate air rearward and thereby propel the cart beyond the wind speed. The energy ultimately comes from the relative motion of the two media---moving air and (approximately) stationary ground. The drivetrain couples these media and converts their velocity difference into thrust, with the wheel--ground contact supplying the required traction (no-slip) constraint and reaction forces.

In practice, the achievable speed ratio is limited by nonidealities---finite propeller efficiency, parasitic aerodynamic drag, rolling resistance and bearing losses, and the finite traction available at the wheel--ground contact (incipient slip). These effects set a maximum useful gear ratio and therefore bound the attainable velocity amplification. A complete engineering treatment would include propulsive efficiency, drag, and traction limits; the idealized analogies below are intended to establish the kinematic and energetic possibility in principle.

\subsection{Upwind Travel}

The same mechanism works in reverse for upwind motion, as depicted in Fig.~\ref{fig:vehicle_operation}(b). Here the wind drives the propeller as a turbine, transferring power through the drivetrain to the wheels, which pull the vehicle against the wind.
The vehicle now extracts energy from the moving air and exerts a force against the ground through its wheel contact. In the ideal limit of rolling without slip on a rigid, stationary surface, the ground itself does no work; it merely supplies the reaction forces needed for the drivetrain to transfer momentum from the air to the vehicle, thereby propelling it.
As emphasized above, propulsion requires the external reaction supplied by the second medium.

\section{Why Not in the Air?}
\label{wind:whynotintheair}

An aircraft, by contrast, operates entirely within a single medium---the air. It must move relative to that air to generate lift and thrust.
A plane in a \SI{100}{km/h} tailwind may achieve a ground speed of \SI{600}{km/h} while its airspeed remains \SI{500}{km/h}, but if its airspeed were reduced to zero, it would lose lift and fall.

A purely wind-driven vehicle of the type analyzed by Gaunaa et al.~\cite{Gaunaa2009}, in contrast, draws power from the relative motion of moving air over (effectively) stationary ground and uses its drivetrain to ``gear up'' this modest velocity difference into a larger vehicle speed.

Because a free-flying aircraft lacks a second medium that can provide an
independent external reaction, it cannot exploit this particular geared
coupling between air and ground.

\section{On Water: Fluid-Mediated Constraints, Slip, and Two Propulsion Scenarios}
\label{sec:water-sail-prop}

Watercraft, like the land vehicles discussed above, couple two media (air and
water) and can exchange momentum between them.  The crucial distinction is that
the ``reaction surface'' is now a \emph{fluid}: hydrodynamic reaction forces are
generated only by imparting momentum to water, which generally produces induced
velocities, a wake, and wake kinetic energy (ultimately dissipated).  Thus the
water-side coupling is not a rigid no-slip traction constraint in the
wheel--ground sense; it is a \emph{fluid-mediated constraint} whose effectiveness
is limited by slip and efficiency.

Throughout this section we return to the ground/water frame \(G\), which for a
boat is the laboratory frame in which the far-field
(undisturbed) water is at rest (neglecting currents). The flow in the vicinity
of the hull/foils/propulsor is generally nonzero due to induced velocities and
wake formation, that is, the
creation of a trailing region of disturbed flow behind the hull or propulsor.
Along the downwind axis, the true wind speed is \(v_w\), the boat speed
through the water is \(v\), and the apparent-wind velocity is
\[
v_a=v_w-v .
\]
Figure~\ref{fig:water-frame} summarizes the kinematic conventions used in this
section: in frame \(G\) the true wind has speed \(v_w\) and the boat moves with
speed \(v\), while in frame \(C\) the apparent-wind velocity is
\(v_a=v_w-v\).

\begin{figure}[tb]
\centering
\begin{tikzpicture}[>=stealth, font=\small, thick]

\node[anchor=east] at (-0.4,0.8) {Frame \(G\):};
\draw[->] (0,0.8) -- (4.2,0.8) node[midway, above] {\(v_w\)};
\draw[->, blue] (0,0.4) -- (2.6,0.4) node[midway, below] {\(v\)};

\node[anchor=east] at (-0.4,-0.6) {Frame \(C\):};
\draw[->, red] (0,-0.6) -- (1.6,-0.6) node[midway, above] {\(v_a\)};
\node[right, red] at (1.7,-0.6) {\(= v_w - v\)};

\end{tikzpicture}
\caption{Minimal kinematic definitions for Sec.~\ref{sec:water-sail-prop}. In the laboratory frame \(G\), the true wind is \(v_w\) and the boat speed is \(v\). In the vehicle frame \(C\), the apparent-wind velocity is \(v_a = v_w - v\). Vectors are aligned at a common origin; for \(v > v_w\), the apparent-wind velocity \(v_a\) reverses direction.}
\label{fig:water-frame}
\end{figure}

\subsection{Scenario 1: Geometric coupling (sails and keels/foils)}

A conventional sailboat acts as an aerodynamic/hydrodynamic ``wedge'' inserted
between air and water: the sail generates an aerodynamic force (typically
lift-dominated with respect to $v_a$), while the keel/foils generate a
hydrodynamic sideforce that prevents lateral drift.

\paragraph*{Against the wind (upwind progress):}
A sailboat does not go straight into the wind; it makes upwind progress by
tacking.  The keel/foil must operate at a finite angle of attack to generate
sideforce, so some leeway and an associated induced-drag penalty are unavoidable
in practice.  Unlike ideal rolling contact, the water-side reaction cannot be
treated as a kinematic constraint that transmits arbitrary lateral force at
zero power cost: generating sideforce requires exchanging momentum with the
water and typically leaves kinetic energy in the wake.

\paragraph*{With the wind (downwind):}
It is common for fast craft (skiffs, catamarans, iceboats) to achieve a downwind
\emph{velocity made good}---the component of velocity directed toward the downwind destination---that exceeds the true wind speed by sailing at
an angle (broad reaching) and jibing; in that regime the apparent wind tends to
increase and rotate forward, keeping the sail in a lifting condition.

By contrast, a conventional sailboat generally cannot sustain a speed exceeding
the wind \emph{directly dead downwind}.  If $v \parallel v_w$
and $v \to v_w$, then $v_a \to 0$ and a passive sail loses
the aerodynamic drive needed to overcome water drag.  To exceed the wind speed
\emph{directly} downwind requires a mechanism that can continue to produce thrust
even when the apparent wind vanishes or reverses---that is, an explicit
``mechanical transformer'' rather than a purely geometric sail--keel coupling.

\subsection{Scenario 2: Mechanical coupling (coupled air and water propellers/turbines)}

To reproduce the Blackbird-style ``two-media gearbox'' on water, one may couple
an air turbine or propeller to a water turbine or propeller via a shaft (and,
if desired, gearing).  Stong and Martin's 1975 account of the Kauffman--Lindahl toy boat
(the ``push--me--pull--you'') provides a particularly clear
demonstration~\cite{Stong1975}.

The key point is a simple force--speed tradeoff. In a one-dimensional
control-volume idealization, if a device processes a mass flow rate \(\dot m\)
and changes the axial fluid speed from \(u_{\rm in}\) to \(u_{\rm out}\), then
\begin{align}
F &= \frac{dp}{dt}
   = \dot m \,(u_{\rm out}-u_{\rm in}), \\
P &= \frac{dE}{dt}
   = \frac{1}{2}\dot m \,(u_{\rm out}^2-u_{\rm in}^2).
\end{align}
Here \(F\) is the axial force exerted on the fluid; the reaction force on the
device has equal magnitude and opposite direction. Thus the force is associated
with the rate of momentum transfer, whereas the power is the rate of energy
transfer. Because kinetic energy increases with the square of speed, even a
modest velocity change in an already fast stream can correspond to a substantial
power transfer.

For a propeller or fan, an ideal actuator-disk estimate makes this scaling more
explicit. If the upstream speed of the fluid relative to the device is \(U\),
the slipstream increment is \(\Delta u\), and the disk area is \(A\), then the
mass flow rate through the disk is
\[
\dot m = \rho A\!\left(U+\frac{\Delta u}{2}\right),
\]
so that the thrust and shaft power transferred to the fluid are
\[
T = \dot m\,\Delta u,
\qquad
P = T\!\left(U+\frac{\Delta u}{2}\right).
\]
In the lightly loaded limit \(\Delta u \ll U\), this reduces to the familiar
scaling
\(
T \approx  {P}/{U}
\).
Thus, for a given shaft power, a device operating in a slower stream can in
principle produce a larger thrust than one operating in a faster stream.
However, efficient operation requires \(\Delta u\) to remain small, which in
turn demands a sufficiently large actuator area; otherwise the wake
kinetic-energy loss rises and the efficiency falls. The corresponding turbine
relations have the same structure, with the sign of the power transfer reversed.

This force--energy intuition must be read
with an important caveat: the achievable forces depend on how much fluid each
device can interact with per unit time, that is, on the mass flow rate
\(\dot m\), which in practice depends strongly on
its effective area (size) and loading.

In a fast stream (air), a turbine can in principle transfer a given shaft power
with a comparatively smaller force because that power is associated with a high
flow speed.  In a slow stream (water), delivering the same shaft power can
correspond to a larger thrust because the characteristic speed is lower.
However, realizing this efficiently requires sufficiently large propeller/turbine
areas: a small, heavily loaded water propeller must produce large slip (a fast
jet), which increases wake losses and reduces thrust per unit power, while a
small air turbine may simply be unable to extract the needed shaft power from
the wind.

Thus the coupled air--water drivetrain can act as a mechanical transformer
between a fast medium and a slow one, but the quantitative outcome depends on
geometry (effective areas) and efficiencies, not on speed alone.  In most
practical implementations the air-side propeller/turbine is much larger than
the water-side one: because air is far less dense than water, extracting (or
delivering) a given shaft power typically requires a much larger swept area in
air, whereas substantial thrust can be produced in water with a comparatively
compact propulsor (often at the cost of increased slip and wake losses if it is
too small).

By reversing the roles---using a \emph{water turbine} (loaded by the boat's
motion through the water) to drive an \emph{air propeller} as a fan---the same
principle applies to downwind travel. Provided the components are sized to
minimize slip penalties, the device can sustain thrust even when the apparent
wind vanishes, enabling downwind travel faster than the wind (in the water
frame).

\subsection{The penalty of slip}

Why is it more difficult to achieve high efficiency (and high speeds) on water than on land?
On land, ideal rolling contact provides a ``hard'' constraint: the instantaneous
contact-point velocity is zero in the ground frame, so the ground can supply
large tangential reaction forces with (ideally) zero power at the contact.

In water, any sustained reaction force requires accelerating a mass flow. This
transfer of kinetic energy to the wake represents an unavoidable energy loss.
Consequently, waterborne faster-than-the-wind devices are
typically more strongly efficiency- and drag-limited than their wheeled
counterparts, even though the underlying Newtonian mechanism is the same: a
transmission mediates momentum exchange between two media in relative motion.

Note that a watercraft relates to Section~\ref{wind:whynotintheair} with an important
qualification. A free-flying aircraft interacts essentially with only one medium
(the air) and therefore lacks access to a second medium in relative motion that
could provide an independent external reaction. A watercraft \emph{does} have such
a second medium (water), but the reaction it provides is mediated by fluid slip and
wake formation rather than by an approximately no-slip traction constraint, as in
ideal wheel--ground contact.

\section{Building Intuition: Mechanical Analogies}

To deepen this understanding, we employ a series of mechanical analogies that
illustrate velocity amplification, energy transfer between coupled media, and
the role of the external reaction constraint introduced above.

\subsection{The Gearbox Analogy}

A gearbox is an excellent first analogy~\cite{veritasium_2021_physics} because it directly mirrors the mechanics of the wind-driven vehicle.
Readers are encouraged to view the demonstration in the video of Ref.~\cite{veritasium_2021_physics} to visualize the effect.
While this mechanical model relies on rigid, no-slip contacts (racks and gears) rather than fluid aerodynamic coupling, it isolates the kinematic principles of velocity amplification, directional transmission, and the essential role of external constraints.
The airflow and ground are modeled as effective racks moving with velocities
\(v_w\) and \(0\), respectively, allowing the use of standard rolling
constraints. The gearbox/vehicle speed is denoted by the same symbol \(v\) as
in the rest of the paper, and the relative upper-rack speed in the vehicle
frame is the apparent-wind velocity \(v_a=v_w-v\).

\subsubsection{Positive Transmission (Downwind)}

In the downwind configuration, the wheels roll on the ground and, through the drivetrain, spin the propeller so that it acts as a fan: it accelerates air rearward (in the ground frame) and thereby produces forward thrust. A useful mechanical analogy is a gearbox operating between two racks: a lower rack rigidly fixed to the ground and an upper rack representing the moving air. The vehicle plays the role of the gearbox casing. By coupling the motion of these two ``racks'' and by virtue of its gear ratio, it converts the relative motion of air and ground into translation of the vehicle over the ground. The essential ingredient is the lower wheel--ground interface: static-friction traction enforces the no-slip rolling constraint and supplies the tangential reaction forces and torque required for the drivetrain to transmit power and exchange momentum with the ground (while doing no work at the contact point in the ideal rolling limit).

To keep the frame changes explicit, we use the vehicle frame only in the next
few lines to fix the sign conventions; the actual speed ratio is then derived
entirely in the ground frame \(G\). Because the gearbox/vehicle may accelerate
in \(G\), the term ``instantaneous rest frame'' means the momentarily comoving
frame at a given instant, used only for local kinematics and sign conventions.
Thus, for the immediate kinematic relations below, we work in the instantaneous
rest frame of the gearbox (``vehicle''), so that all gear centers are at rest.
In this frame the upper rack (``wind'') moves with speed $v_a=v_w-v$ relative to the
gearbox, where $v_w$ denotes the wind speed in the ground frame  \(G\) (used in Section~\ref{sec:gearbox-ground-frame} below),
and the lower rack (``ground'') moves backward with speed $v$; the
same symbol $v$ denotes the ground speed of the gearbox in \(G\).

\begin{figure*}[htbp]
\centering
\begin{tabular}{ccc}
\begin{tikzpicture}[line cap=round,line join=round,>=Stealth,font=\small]
  \useasboundingbox (-3.2,-0.55) rectangle (5.2,6.95);

  \def\rA{1.45}   
  \def\rB{1.75}   
  \def\rP{0.78}   
  \def\rI{0.55}   

  \pgfmathsetmacro{\rAa}{0.82*\rA}
  \pgfmathsetmacro{\rBb}{0.82*\rB}
  \pgfmathsetmacro{\rPp}{0.68*\rP}

  \pgfmathsetmacro{\yOthree}{\rB}             
  \pgfmathsetmacro{\yOone}{\rB+\rP+\rA}       
  \pgfmathsetmacro{\yTop}{\rB+\rP+2*\rA}      

  \coordinate (O3) at (0,\yOthree);
  \coordinate (O1) at (0,\yOone);

  \draw[thick] (-3.0,\yTop) -- (4.9,\yTop);
  \node[above] at (-2,\yTop+0.2) {upper rack};
  \draw[->,thick] (2.8,\yTop) -- (4.35,\yTop);
  \node[above] at (3.58,\yTop+0.2) {$v_w$};

  \draw[dashed] (-2.25,\yOone) -- (2.25,\yOone);
  \draw[dashed] (-2.25,\yOthree) -- (2.25,\yOthree);
  \draw[dashed] (-2.25,\yOthree) -- (-2.25,\yOone);
  \draw[dashed] (2.25,\yOthree) -- (2.25,\yOone);

  \draw[thick] (O1) circle (\rA);
  \fill (O1) circle (1.5pt);

  \draw[thick] (O3) circle (\rB);
  \draw[thick] (O3) circle (\rP);
  \fill (O3) circle (1.5pt);

 \draw[->,thick] ($(O1)+(-30:\rAa)$)
  arc[start angle=-30,end angle=50,radius=\rAa];
  \draw[->,thick] ($(O3)+(40:\rBb)$)
    arc[start angle=40,end angle=-40,radius=\rBb];
  \draw[->,thick] ($(O3)+(25:\rPp)$)
    arc[start angle=25,end angle=-55,radius=\rPp];

  \node at (0,\yOone+0.28) {$v_1=r_1\omega_1$};
  \node at (0,\yOthree-1.05) {$v_2=r_2\omega_2=-v_1$};
  \node[below right] at (1.4,0.7) {$v_3=r_3\omega_3=r_3\omega_2$};

  \draw[thick] (-2.9,0) -- (4.9,0);
  \node[below] at (-2.2,-0.2) {ground};
  \foreach \x in {-1.3,-0.9,...,4.7}
    \draw (\x,0) -- ++(-0.18,-0.24);

  \draw[->,thick] (2.5,2.55) -- (4.45,2.55);
  \draw[->,thick] (3.5,2.55) -- (4.75,2.55);
  \node[above] at (3.95,2.55) {$v$};
\end{tikzpicture}
&$\qquad$&
\begin{tikzpicture}[line cap=round,line join=round,>=Stealth,font=\small]
  \useasboundingbox (-3.2,-0.55) rectangle (5.2,6.95);

  \def\rA{1.45}   
  \def\rB{1.75}   
  \def\rP{0.78}   

  \pgfmathsetmacro{\rAa}{0.82*\rA}
  \pgfmathsetmacro{\rBb}{0.82*\rB}
  \pgfmathsetmacro{\rPp}{0.68*\rP}

  \pgfmathsetmacro{\yOone}{\rA}                  
  \pgfmathsetmacro{\yOthree}{2*\rA+\rP}          
  \pgfmathsetmacro{\yTop}{\yOthree+\rB}          

    \draw[thick] (-3.0,\yTop) -- (4.9,\yTop);
    \node[above] at (-2,\yTop+0.2) {upper rack};
    \draw[->,thick] (2.8,\yTop) -- (4.35,\yTop);
    \node[above] at (3.58,\yTop+0.2) {$v_w$};

  \begin{scope}[xshift=2.45cm]

    \coordinate (O1) at (0,\yOone);
    \coordinate (O3) at (0,\yOthree);

    \draw[dashed] (-2.25,\yOone) -- (2.25,\yOone);
    \draw[dashed] (-2.25,\yOthree) -- (2.25,\yOthree);
    \draw[dashed] (-2.25,\yOone) -- (-2.25,\yOthree);
    \draw[dashed] (2.25,\yOone) -- (2.25,\yOthree);

    \draw[thick] (O3) circle (\rB);
    \draw[thick] (O3) circle (\rP);
    \fill (O3) circle (1.5pt);

    \draw[thick] (O1) circle (\rA);
    \fill (O1) circle (1.5pt);

    \draw[->,thick] ($(O3)+(28:\rBb)$)
  arc[start angle=28,end angle=-40,radius=\rBb];
    \draw[->,thick] ($(O3)+(25:\rPp)$)
      arc[start angle=25,end angle=-55,radius=\rPp];
 \draw[->,thick] ($(O1)+(-45:\rAa)$)
  arc[start angle=-45,end angle=60,radius=\rAa];

    \node at (0,\yOthree+1.35) {$v_3=r_3\omega_3$};
    \node at (0,\yOthree+1) {$v_2=r_2\omega_2=r_2\omega_3$};
    \node at (0,\yOone-0.42) {$v_1=r_1\omega_1$};

  \end{scope}

    \draw[thick] (-2.9,0) -- (4.9,0);
    \node[below] at (-2.2,-0.2) {ground};
    \foreach \x in {-1.3,-0.9,...,4.7}
      \draw (\x,0) -- ++(-0.18,-0.24);

  \draw[->,thick] (-0.15,2.55) -- (-2.10,2.55);
  \draw[->,thick] (-0.10,2.55) -- (-1.80,2.55);
  \node[above] at (-1.15,2.55) {$v$};

\end{tikzpicture}
\\
$\quad$
\\
(a) Positive Transmission && (b) Negative Transmission
\end{tabular}
\caption{Schematic of the gearbox analogy in the laboratory frame \(G\).
The lower line represents the ground, which is stationary in this frame, and
the arrow \(v\) denotes the velocity of the gearbox/vehicle relative to the
ground.  Curved arrows indicate the sense of rotation, and the labels \(v_1\),
\(v_2\), and \(v_3\) denote local tangential speeds as defined in the text.
The apparent sign mismatch at a rack--gear contact is resolved by the
translation of the gearbox: because the gear center moves with the gearbox, the
no-slip condition in the laboratory frame involves the \emph{absolute}
contact-point velocity, obtained by adding the translational velocity \(v\)
to the local tangential velocity.  (a) Positive transmission, or downwind
configuration: the upper rack acts on the free upper gear, which drives the
pinion and the lower gear rigidly attached to it.  The resulting velocity ratio
is \(v/v_w=r_3/(r_3-r_2)\).  (b) Negative transmission, or upwind
configuration: the gearbox is the top--bottom inverted counterpart of
(a).  The gear of radius \(r_3\), still rigidly attached to the pinion of
radius \(r_2\), now contacts the upper rack, while the free gear of radius
\(r_1\) contacts the ground.
This gives \(v/v_w=-r_2/(r_3-r_2)\).
For the particular radii drawn in panel (b), the motion is upwind because
\(v/v_w<0\); however, it is faster than the wind only when
\(r_2<r_3<2r_2\), equivalently \(1/2<r_2/r_3<1\).}
\label{fig:gearboxes}
\end{figure*}

We can model this using a system of racks and gears, as shown in Fig.~\ref{fig:gearboxes}(a). Consider a mechanical system with:
\begin{enumerate}
\item An upper rack representing the wind, moving with velocity $v_a$ in the vehicle frame (equivalently, speed $v_w$ in $G$), with no slippage between the rack and the upper gear.
\item An upper gear of radius $r_1$ in contact with the upper rack, rotating with angular velocity $\omega_1$.
\item A pinion of radius $r_2$ meshed with the gear, rotating with $\omega_2$.
\item A lower gear of radius $r_3$ rigidly attached to the pinion, rotating with $\omega_3 = \omega_2$.
\item A lower rack representing the ground, moving backward with speed $v$ in the vehicle frame, with no slippage at the lower gear. It is stationary in the laboratory frame $G$.
\end{enumerate}

We take rightward motion as positive along the horizontal axis.
With counter-clockwise rotation taken as positive, the kinematic relations in
the instantaneous rest frame of the gearbox are
\begin{align}
v_a &= -v_1,  \label{eq:pos_rigid}\\
v_1 &= r_1 \omega_1,\\
v_2 &= r_2 \omega_2 = -v_1, \label{eq:pos_mesh}\\
v_3 &= r_3 \omega_3 = r_3 \omega_2.
\end{align}
All tangential velocities \(v_k = r_k \omega_k\) are understood as signed horizontal
components evaluated at the respective contact points. With the convention that
counter-clockwise angular velocity is positive, the sign of \(v_k\) follows from the
local direction of motion of the gear tooth at the point of contact.

The two negative signs---one from the top-side rack contact in
Eq.~\eqref{eq:pos_rigid} and one from the gear--gear mesh in
Eq.~\eqref{eq:pos_mesh}---cancel, so the tangential velocities at the upper and
lower rack contacts point in the \emph{same} direction. Thus the gearbox
transmits motion from the upper rack to the lower rack without reversing its
direction: this is what we call \emph{positive transmission}.

We now switch back to the ground frame \(G\), because the no-slip condition at
the lower rack and the comparison with the wind speed are most transparent in
that frame. The actual velocity and force ratios between the racks, and hence
between the ``wind'' and the vehicle over the ground, are therefore obtained by
writing the no-slip conditions in \(G\).

\subsubsection{Positive Transmission in the ground frame}
\label{sec:gearbox-ground-frame}

We now write the kinematics entirely in the ground frame. Let the upper rack
(``wind'') move to the right with speed $v_w$, and let the gearbox/vehicle
translate to the right with speed $v$. The center of the upper gear therefore
moves to the right at speed $v$ in this frame. Denote by
$
v_1 = r_1 \omega_1
$
the tangential speed of the upper gear at its rim, measured in the gear's own
rest frame, with counter-clockwise rotation taken as positive.

At the top contact point between the upper gear and the rack, a positive
(counter-clockwise) \(\omega_1\) produces a leftward tangential velocity of the
tooth, that is, a negative horizontal component opposite to the rack motion.
Thus, in the ground frame, the horizontal velocity of that contact point is
the sum of the translational and tangential contributions, \(v-v_1\).
The no-slip condition at the upper rack--gear interface then requires the tooth
velocity to equal the rack velocity in the ground frame,
\begin{equation}
v_w = v - v_1.
\label{eq:upper-noslip-ground}
\end{equation}
This is the ground-frame version of Eq.~\eqref{eq:pos_rigid}, $v_a = -v_1$, originally written in the instantaneous rest frame of the gearbox, since $v_a=v_w-v$.

The remaining kinematic relations can also be written in the ground frame.
Let again  $
v_2 = r_2 \omega_2$ and $ v_3 = r_3 \omega_3
$
be the tangential speeds of the pinion and the lower gear, respectively,
measured in their local rest frames. At the mesh between the upper gear and the
pinion, the centers have the same translational speed $v$, so this cancels
out, and no slip requires equal and opposite tangential velocities along the
line of contact,
\begin{equation}
v_2 = -\,v_1,
\label{eq:mesh-v2}
\end{equation}
reflecting the opposite tangential directions at the pitch point of two externally meshing gears.
Because the pinion and lower gear are rigidly fixed to the same shaft,  $\omega_3 = \omega_2$, and  therefore,
\begin{equation}
v_3 = r_3 \omega_3 = \frac{r_3}{r_2}\,v_2.
\label{eq:v3-from-v2}
\end{equation}
Combining Eqs.~\eqref{eq:mesh-v2} and~\eqref{eq:v3-from-v2} gives
\begin{equation}
v_3 = \frac{r_3}{r_2} v_2
    = -\,\frac{r_3}{r_2} v_1.
\label{eq:v3-v1}
\end{equation}

Thus \(v_3\) is negative whenever \(v_1\) is positive, implying that the lower
gear rotates clockwise in this configuration.
At the lower contact, between the lower gear and the (stationary) lower rack
representing the ground, the gear center again moves to the right with speed
$v$.
The contact-point velocity is the sum of the translational velocity of the gear
center, \(v\), and the tangential velocity at the rim, \(v_3 = r_3 \omega_3\),
taken with its sign.
The no-slip condition in the ground frame (where the lower rack is at rest)
requires
\begin{equation}
0 = v + v_3
\text{ or }
v = -\,v_3.
\label{eq:lower-noslip-ground}
\end{equation}

Using \eqref{eq:v3-v1} in \eqref{eq:lower-noslip-ground} yields
\begin{equation}
v = -v_3 = \frac{r_3}{r_2}\,v_1.
\label{eq:v-v1}
\end{equation}
On the other hand, from the upper rack--gear no-slip condition
\eqref{eq:upper-noslip-ground} we have
\begin{equation}
v_1 = v - v_w.
\label{eq:v1-from-vw-v}
\end{equation}
Substituting \eqref{eq:v1-from-vw-v} into \eqref{eq:v-v1} gives
\[
v = \frac{r_3}{r_2}\,(v - v_w),
\]
and, solving for $ {v}/{v_w}$  yields
\begin{equation}
 v = \frac{r_3}{r_3 - r_2}\, {v_w}.
\label{eq:v-vw-ratio}
\end{equation}
The gearbox (and hence the vehicle) moves faster than the upper rack
(``wind'') by the velocity-amplification factor
\begin{equation}
 \frac{v}{v_w} = \frac{r_3}{r_3 - r_2},
\label{eq:amp-v-vw-ratio}
\end{equation}
which is greater than unity whenever $r_3 > r_2$.
The expression requires \(r_3 > r_2\) to yield a positive forward velocity \(v\).
This amplification arises purely from the geometric constraints imposed by
the two no-slip conditions and the gear coupling, independent of any
dynamical assumptions.
Notably, this kinematic
velocity ratio is independent of the upper gear radius $r_1$.
Substituting $v_w= v_a + v$ into Eq.~\eqref{eq:v-vw-ratio} and using the vehicle-frame relation Eq.~\eqref{eq:pos_rigid} recovers Eq.~\eqref{eq:v-v1}, confirming consistency between the two frames.

\begin{figure*}[t]
\centering

\begin{minipage}{0.48\textwidth}
\centering
\begin{tikzpicture}
\begin{axis}[
    width=\linewidth,
    height=5.6cm,
    xmin=0, xmax=0.95,
    ymin=1, ymax=8,
    xlabel={$x=r_2/r_3$},
    ylabel={$v/v_w$},
    axis y line*=left,
    axis x line*=bottom,
    ylabel style={text=blue},
    yticklabel style={text=blue},
    xtick={0,0.2,0.4,0.6,0.8},
    ytick={1,2,4,6,8},
    grid=both,
    grid style={line width=.1pt, draw=gray!25},
    major grid style={line width=.2pt, draw=gray!45},
    legend style={draw=none, fill=none, font=\small, at={(0.05,0.14)}, anchor=south west}
]
\addplot[blue, very thick, domain=0:0.95, samples=300] {1/(1-x)};
\addlegendentry{$\displaystyle \frac{v}{v_w}=\frac{1}{1-x}$}
\end{axis}

\begin{axis}[
    width=\linewidth,
    height=5.6cm,
    xmin=0, xmax=0.95,
    ymin=0, ymax=1.05,
    axis y line*=right,
    axis x line=none,
    xtick=\empty,
    ylabel={$F/F_w$},
    ylabel style={text=red},
    yticklabel style={text=red},
    ytick={0,0.2,0.4,0.6,0.8,1.0},
    legend style={draw=none, fill=none, font=\small, at={(0.4,0.7)}, anchor=west}
]
\addplot[red, very thick, domain=0:0.95, samples=2] {1-x};
\addlegendentry{$\displaystyle \frac{F}{F_w}=1-x$}
\end{axis}
\end{tikzpicture}

\vspace{1mm}
\textbf{(a)} Gearbox tradeoff
\end{minipage}
\hfill
\begin{minipage}{0.48\textwidth}
\centering
\begin{tikzpicture}
\begin{axis}[
    width=\linewidth,
    height=5.6cm,
    xmin=0, xmax=2,
    ymin=-1.1, ymax=1.1,
    xlabel={$v/v_w$},
    ylabel={$v_a/v_w$},
    xtick={0,0.5,1,1.5,2},
    ytick={-1,-0.5,0,0.5,1},
    grid=both,
    grid style={line width=.1pt, draw=gray!25},
    major grid style={line width=.2pt, draw=gray!45},
]
\addplot[black, very thick, domain=0:2, samples=2] {1-x};
\addplot[dashed, gray] coordinates {(1,-1.1) (1,1.1)};
\addplot[dashed, gray] coordinates {(0,0) (2,0)};

\node[font=\small, anchor=south west] at (axis cs:0.35,0.7) {tailwind in $C$};
\node[font=\small, anchor=south west] at (axis cs:1.05,-0.95) {headwind in $C$};
\node[font=\small, anchor=south] at (axis cs:1.2,0.04) {crossover};
\end{axis}
\end{tikzpicture}

\vspace{1mm}
\textbf{(b)} Apparent wind in the vehicle frame
\end{minipage}

\caption{Dimensionless plots of two key relations used in the text.
(a) In the gearbox analogy, with $x=r_2/r_3$, the ideal speed amplification
$\frac{v}{v_w}=\frac{1}{1-x}$ increases as $x\to1^{-}$, while the transmitted
force ratio $\frac{F}{F_w}=1-x$ decreases correspondingly. This illustrates
the force--speed tradeoff of the lossless transmission.
(b) In the vehicle frame \(C\), the apparent wind satisfies
$\frac{v_a}{v_w}=1-\frac{v}{v_w}$. It is positive for $v<v_w$, vanishes at the
crossover $v=v_w$, and becomes negative for $v>v_w$, meaning that the air then
approaches from ahead in the vehicle frame.}
\label{fig:dimensionless_plots}
\end{figure*}

Forces are treated as signed quantities along the horizontal axis,
consistent with the velocity conventions introduced above.  The ideal gearbox
acts as a lossless mechanical transformer between the moving upper rack and
the translating gearbox/vehicle.  Thus the useful mechanical power associated
with the upper rack force and the vehicle motion satisfies
\[
F_w v_w = F v ,
\]
where \(F_w\) is the tangential force transmitted at the upper rack--gear
contact and \(F\) is the corresponding effective horizontal reaction associated
with the lower gear--ground constraint.  This equation should not be read as
work done by the ground at the contact point: in the laboratory frame the
ideal wheel--ground contact point has zero velocity, so the ground contact
itself does no work.
Combining this with Eq.~\eqref{eq:v-vw-ratio} yields
\begin{equation}
F = \frac{v_w}{v} \,  F_w
    = \frac{r_3 - r_2}{r_3}\, F_w.
\end{equation}
Thus the larger vehicle speed is accompanied by a proportionally reduced
effective horizontal force: an increase in speed is obtained at the cost of a
corresponding decrease in transmitted force, exactly as in a conventional gear
train.

The dimensionless behavior of these relations is shown in Fig.~\ref{fig:dimensionless_plots}(a): as the idealized speed amplification grows, the transmitted force decreases proportionally.

Note that the velocity ratio \eqref{eq:v-vw-ratio}
becomes singular in the limit $r_2 \to r_3$. This does not imply that the
vehicle speed literally diverges; rather, it signals that our idealized
no-slip model ceases to have a consistent solution for $r_2 = r_3$ and
prescribed $v_w \neq 0$.

Indeed, before solving for $v$ we had, from the upper no-slip condition
\eqref{eq:upper-noslip-ground} and the gear relations \eqref{eq:v-v1},
\[
v_w = \left(\frac{r_3}{r_2} - 1\right)v_1
     = \frac{r_3 - r_2}{r_2}\,v_1.
\]
For $r_2 = r_3$ this reduces to  $v_w = 0$,
so the only way to satisfy all no-slip constraints with $r_2 = r_3$ is to
have zero relative motion between the racks. Thus, if the upper rack is
prescribed to move at a nonzero speed $v_w$, at least one of the contacts
must slip or the mechanism must deform, and our ideal rigid, no-slip model
breaks down exactly at $r_2 = r_3$. The apparent ``singularity'' in
\eqref{eq:v-vw-ratio} therefore marks a kinematic incompatibility, not a
physical divergence of $v$.

An intuitive but somewhat ``impromptu'' kitchen-style elementary demonstration for the degenerate case $r_2 = r_3$ uses two apples or two rolling pins, one stacked on the other, between a table (the ``ground'') and a wooden or plastic cutting board (the ``upper rack'').
If the two rollers are constrained to translate together as a unit (their centers share the same horizontal velocity), and all three contacts (ground--lower roller, roller--roller, upper roller--board) are rolling without slip, then the cutting board cannot move at all relative to the table. Nevertheless, the ``apple (or pin) train'' itself can move arbitrarily relative to both, since the internal rolling still admits a free translational speed.

As emphasized above, this mechanism works only because the chassis is externally
constrained by the wheel--ground contact. Without that traction-mediated
coupling, the assembly would simply drift with the air, and no sustained
velocity amplification relative to the environment would result.

\subsubsection{Negative Transmission (Upwind)}

For upwind travel, the \emph{energetic} roles are reversed: the wind now drives
the propeller as a turbine, and the drivetrain transmits this power to the
wheels, which push against the ground to move the vehicle into the wind.
Kinematically, however, the appropriate gearbox analogue is not obtained by
inserting an idler gear into the positive-transmission train.  An inserted
idler reverses the sense of rotation in an isolated gear train, but in the
present rack--gear--rack topology it also changes the closure imposed by the
two no-slip rack constraints.  The resulting motion is still downwind.

The clean negative-transmission analogue is instead the top--bottom inverted
version of the positive-transmission gearbox, shown in
Fig.~\ref{fig:gearboxes}(b).  In this configuration the gear of radius \(r_3\),
still rigidly attached to the pinion of radius \(r_2\), contacts the upper rack
representing the wind, while the free gear of radius \(r_1\) contacts the lower
rack representing the ground.  The upper and lower racks themselves are not
interchanged: the upper rack still moves with velocity \(v_w\) in the ground
frame \(G\), and the lower rack remains stationary.

We again take rightward motion as positive and counter-clockwise angular
velocity as positive.  Let \(\omega_3\) denote the angular velocity of the
upper gear of radius \(r_3\), which is rigidly attached to the pinion of radius
\(r_2\), and let \(\omega_1\) denote the angular velocity of the lower free
gear of radius \(r_1\).  The gearbox/vehicle translates with velocity \(v\)
in the ground frame.

At the upper rack--gear contact, a positive \(\omega_3\) gives a leftward
tangential velocity at the top of the gear.  Therefore the absolute velocity
of the upper contact point is
\[
v-r_3\omega_3 .
\]
No slip at the upper rack requires this to equal the rack velocity \(v_w\):
\begin{equation}
v_w = v-r_3\omega_3 .
\label{eq:upwind-upper-noslip}
\end{equation}

At the mesh between the pinion of radius \(r_2\) and the lower free gear of
radius \(r_1\), the two gears mesh externally, so their tangential velocities
at the pitch point are equal and opposite:
\begin{equation}
r_2\omega_3 = -\,r_1\omega_1 .
\label{eq:upwind-mesh}
\end{equation}

At the lower gear--ground contact, the lower rack is stationary in the ground
frame.  The center of the lower gear moves with velocity \(v\), while the
tangential velocity at the lower contact is \(r_1\omega_1\).  The no-slip
condition is therefore
\begin{equation}
0 = v+r_1\omega_1 .
\label{eq:upwind-lower-noslip}
\end{equation}

Equations~\eqref{eq:upwind-mesh} and~\eqref{eq:upwind-lower-noslip} imply
\[
r_1\omega_1=-v,
\qquad
r_2\omega_3=v,
\]
and hence
\[
\omega_3=\frac{v}{r_2}.
\]
Substitution into the upper no-slip condition
\eqref{eq:upwind-upper-noslip} gives
\[
v_w
 =
v-r_3\frac{v}{r_2}
 =
v\left(1-\frac{r_3}{r_2}\right)
 =
v\,\frac{r_2-r_3}{r_2}.
\]
Solving for \(v/v_w\), one obtains
\begin{equation}
\frac{v}{v_w}
=
-\frac{r_2}{r_3-r_2}
\qquad\text{(negative transmission).}
\label{eq:upwind-v-vw-ratio}
\end{equation}
Thus, for \(r_3>r_2\), the velocity \(v\) is negative when \(v_w\) is
positive: the gearbox, and hence the vehicle, moves upwind.

The magnitude of this negative-transmission ratio is not the same as in the
positive-transmission case.  To compare the two orientations without
overloading the symbol \(v\), let \(v_{\rm down}\) and \(v_{\rm up}\) denote
the signed ground-frame velocities of the positive-transmission and
top--bottom inverted configurations, respectively, for the same \(v_w\) and
the same gear radii.  The positive configuration gave
\[
D_{\rm down}
:=
\frac{v_{\rm down}}{v_w}
=
\frac{r_3}{r_3-r_2},
\]
whereas the top--bottom inverted configuration gives
\[
D_{\rm up}
:=
\frac{v_{\rm up}}{v_w}
=
-\frac{r_2}{r_3-r_2}.
\]
These two signed ratios obey the simple identity
\begin{align}
&D_{\rm down} + D_{\rm up}
=
\frac{r_3}{r_3-r_2}
-
\frac{r_2}{r_3-r_2}
=1,
\nonumber\\
&\text{or }\qquad
D_{\rm up}=1-D_{\rm down}.
\label{eq:upwind-downwind-symmetry}
\end{align}
Equivalently, the signed ground-frame velocities of the downwind and upwind
orientations satisfy
\[
v_{\rm down}+v_{\rm up}=v_w .
\]
This relation is purely kinematic; it follows from the two rack constraints
and from using the same gear radii in the two top--bottom inverted
orientations.

It is useful to separate two different questions: first, whether the inverted
gearbox moves upwind at all, and second, whether its upwind speed exceeds the
wind speed.  We take \(v_w>0\), so upwind motion corresponds to \(v_{\rm up}<0\), or
equivalently \(D_{\rm up}<0\).  Since the gear radii are positive,
\[
D_{\rm up}
=
-\frac{r_2}{r_3-r_2}
\]
is negative precisely when
\[
r_3-r_2>0.
\]
Thus the inverted gearbox moves upwind exactly when
\[
r_3>r_2 .
\]

The stronger condition for faster-than-the-wind upwind motion is not merely
\(D_{\rm up}<0\), but
\[
|D_{\rm up}|>1 .
\]
Once the upwind branch \(r_3>r_2\) has been selected, the denominator
\(r_3-r_2\) is positive, and therefore
\[
|D_{\rm up}|
=
\frac{r_2}{r_3-r_2}.
\]
Hence faster-than-the-wind upwind motion requires
\[
\frac{r_2}{r_3-r_2}>1,
\]
which is equivalent to
\[
r_3<2r_2 .
\]
Combining the direction condition \(r_3>r_2\) with the speed condition
\(r_3<2r_2\), we obtain
\begin{equation}
r_2<r_3<2r_2 .
\label{eq:upwind-amplification-condition}
\end{equation}
Thus the top--bottom inverted gearbox moves upwind whenever \(r_3>r_2\), but
it moves upwind faster than the wind only in the narrower interval
\(r_2<r_3<2r_2\).

The same result can be read directly from the complementarity relation
\eqref{eq:upwind-downwind-symmetry}.  Since
\[
D_{\rm up}=1-D_{\rm down},
\]
the inverted gearbox is faster than the wind upwind exactly when
\[
D_{\rm up}<-1,
\]
or equivalently
\[
D_{\rm down}>2.
\]
Thus an upwind speed exceeding the wind corresponds to the associated
positive-transmission orientation having a downwind speed ratio larger than
two.

In the ideal, lossless limit, conservation of mechanical power again gives
\[
F_w v_w = F v ,
\]
with forces understood as signed horizontal quantities.  Since \(v\) has the
opposite sign to \(v_w\) in the upwind configuration, the corresponding force
at the lower rack has the opposite sign to \(F_w\).  In magnitude,
Eq.~\eqref{eq:upwind-v-vw-ratio} gives
\begin{equation}
|F|
=
\frac{|v_w|}{|v|}\,|F_w|
=
\frac{r_3-r_2}{r_2}\,|F_w|.
\label{eq:upwind-force-ratio}
\end{equation}
Thus upwind speed amplification, when present, is again obtained at the cost of
a corresponding reduction in transmitted force, but with the amplification
factor \(r_2/(r_3-r_2)\) rather than \(r_3/(r_3-r_2)\).

In the physical vehicle, the wind supplies input power to the propeller, which
acts as a turbine.  The drivetrain transmits this power to the wheels, and the
wheel--ground contact supplies the external reaction needed for the vehicle to
move against the wind.  As in the downwind case, the ideal contact point with
the ground does no work in the rolling limit; its role is to impose the
constraint and provide the necessary horizontal reaction.

\begin{figure*}[htbp]
\centering
\begin{tabular}{ccc}
\begin{tikzpicture}[
    line cap=round,
    line join=round,
    >=Stealth,
    font=\normalsize
  ]
  \useasboundingbox (-4.2,0.6) rectangle (4.5,7.2);

  \begin{scope}[yshift=1.5cm]

    \coordinate (P)   at (0,0);      
    \coordinate (T)   at (0,4.6);    
    \coordinate (Tf)  at (2.15,4.0); 

    \draw[thick] (-4.0,0) -- (4.1,0);

    \draw[thick] (-0.18,-0.22) -- (P) -- (0.18,-0.22) -- cycle;
    \fill (0,0) circle (2pt);

    \draw[thick] (P) -- (T);

    \draw[thick] (-0.55,4.6) -- (0.55,4.6);

    \draw[dashed] (P) -- (Tf);
    \draw[dashed] ($(Tf)+(-0.42,0.18)$) -- ($(Tf)+(0.42,-0.18)$);

    \draw[thick, fill=white] (-1.05,0.95) rectangle (1.05,2.35);
    \node at (0,1.65) {sail};

    \draw[->,thick] ($(P)+(87:3.5)$) arc[start angle=87,end angle=65,radius=3.5];
    \draw[->,thick] ($(P)+(87:3.5)$) arc[start angle=87,end angle=70,radius=3.5];

    \node at (-3.0,4.2) {wind};
    \foreach \y in {3.0,2.1,1.2}{
      \draw[->,thick,decorate,decoration={snake,amplitude=2pt,segment length=18pt}]
        (-3.75,\y) -- (-1.95,\y);
    }

    \node at ($(P)+(0,5.0)$) {platform};
    \node at ($(P)+(0,-0.55)$) {pivot};

  \end{scope}
\end{tikzpicture}
&&
\begin{tikzpicture}[
    line cap=round,
    line join=round,
    >=Stealth,
    font=\normalsize
  ]
  \useasboundingbox (-4.2,-0.4) rectangle (4.5,8.2);

  \def\thCur{15}   
  \def\thFut{58}   

  \node at (-0.55,4.55) {wind};
  \draw[->,thick,decorate,decoration={snake,amplitude=2pt,segment length=18pt}]
    (-4.0,4.15) -- (-2.85,3.25);
  \draw[->,thick,decorate,decoration={snake,amplitude=2pt,segment length=18pt}]
    (-3.55,4.75) -- (-2.35,3.85);
  \draw[->,thick,decorate,decoration={snake,amplitude=2pt,segment length=18pt}]
    (-3.1,5.15) -- (-1.9,4.25);

  \begin{scope}[yshift=1cm]

    \coordinate (P)       at (1.05,1.25);                    
    \coordinate (Sattach) at ($(P)+({180+\thCur}:2.45)$);   
    \coordinate (R)       at ($(P)+(\thCur:2.95)$);         
    \coordinate (Rf)      at ($(P)+(\thFut:3.25)$);         
    \coordinate (Rb)      at ($(P)+({180+\thFut}:2.15)$);   

    \draw[thick] (Sattach) -- (R);

    \draw[dashed] (Rb) -- (Rf);

    \draw[thick] ($(P)+(-0.18,-0.22)$) -- (P) -- ($(P)+(0.18,-0.22)$) -- cycle;
    \fill (P) circle (2pt);

    \draw[thick] ($(R)+({\thCur+90}:0.55)$) -- ($(R)+({\thCur-90}:0.55)$);

    \draw[dashed] ($(Rf)+({\thFut+90}:0.55)$) -- ($(Rf)+({\thFut-90}:0.55)$);

    \node[
      draw,
      thick,
      fill=white,
      minimum width=1.20cm,
      minimum height=2.10cm,
      rotate=\thCur,
      anchor=east,
      inner sep=0pt
    ] at (Sattach) {sail};

    \draw[->,thick] ($(P)+(200:1.15)$) arc[start angle=200,end angle=234,radius=1.15];
    \draw[->,thick] ($(P)+(200:1.15)$) arc[start angle=200,end angle=224,radius=1.15];
    \draw[->,thick] ($(P)+(18:2.0)$)   arc[start angle=18,end angle=56,radius=2.0];
    \draw[->,thick] ($(P)+(18:2.0)$)   arc[start angle=18,end angle=46,radius=2.0];

    \node at ($(Rf)+(1.0,0.10)$) {platform};
    \node at ($(P)+(0.20,-0.55)$) {pivot};

  \end{scope}
\end{tikzpicture}
\\
(a) Downwind Analogy && (b) Upwind Analogy
\end{tabular}
\caption{Schematic of the lever analogy. (a) For downwind motion, the pivot is at one end. The wind pushes the sail (at distance $d_s$), causing the far end, a distance \(L\) from the pivot, to move faster than the wind. (b) For upwind motion, the pivot lies between the sail and the observation point, causing the far end to move in the opposite direction of the wind's push.}
\label{fig:lever-downupwind}
\end{figure*}

\subsection{The Lever Analogy}

A lever provides a second, complementary analogy for how speed can be amplified
through lever geometry and a fixed external support. Here geometry replaces
gearing, but the essential ideas---velocity amplification, possible reversal of
direction, and the need for an external constraint that supplies reaction
forces and torque---are the same. As before, the lever represents the
\emph{vehicle itself}, mediating between the wind and the ground.
The kinematic relations are formally analogous to those of a lever,
with the radii \(r_2\) and \(r_3\) playing the role of effective lever arms.

However, it must be noted that a physical lever allows only transient faster-than-the-wind motion, limited by the length of its stroke. The wind-driven vehicle, however, operates as a \emph{continuous} rotary lever, maintaining this kinematic transformation indefinitely as it rolls.

\subsubsection{With the Wind}

Consider a rigid lever of total length $L$, pivoted at one end and firmly clamped to the bench (fixed pivot), as shown in Fig.~\ref{fig:lever-downupwind}(a). A sail is attached partway along the lever at distance $d_s$ from the pivot, and we observe the motion of the far end. When the wind pushes the sail, the lever rotates about its pivot. By geometry, the end of the lever moves faster than the point where the sail is attached.

Let the wind blow at speed $v_w$, and let the sail move with velocity $v_s = \alpha v_w$, where $0 < \alpha < 1$ accounts for aerodynamic loss. The angular velocity of the lever is
\begin{equation}
\omega = \frac{v_s}{d_s}.
\end{equation}
The velocity of the distal end is then
\begin{equation}
v_d = \omega L = \alpha \left(\frac{L}{d_s}\right) v_w.
\end{equation}
Hence, the velocity amplification factor is simply $L/d_s$, independent of the absolute dimensions of the lever. If $L/d_s > 1/\alpha$, the lever's tip moves faster than the wind ($v_d > v_w$).
This ratio plays the same role as the velocity amplification factor
$r_3/(r_3 - r_2)$ in the gearbox analogy.

In the ideal lossless limit, the lever does not save energy or power; rather,
it trades force for distance (or, equivalently, force for speed). Thus the
amplification of speed comes with a proportional reduction in force:
\begin{equation}
F_d = F_w \frac{d_s}{L},
\end{equation}
where $F_w$ is the wind force on the sail and $F_d$ is the force at the lever tip. The pivot acts as an external constraint (a fixed support) that supplies the required reaction forces and torque, enabling momentum exchange between the wind-driven part of the lever and the ground.

\subsubsection{Against the Wind}

For upwind motion, we move the pivot between the sail and the far end of the
lever, as illustrated in Fig.~\ref{fig:lever-downupwind}(b). When the wind
pushes on the sail, the distal end moves in the opposite direction---against
the wind. The reversal is a consequence of the lever geometry itself: the sail
and the observed endpoint lie on opposite sides of the fixed pivot. The same
lever-arm ratio still controls the magnitude of the velocity amplification.

Let the pivot be at distance $d_p$ from the sail, and let the far end of the lever be $L - d_p$ away on the opposite side. If the wind drives the sail with velocity $v_s = \alpha v_w$ (with the same convention $0 < \alpha < 1$ for the effective coupling factor), the lever rotates with angular velocity
\begin{equation}
\omega = \frac{v_s}{d_p}.
\end{equation}
The velocity at the opposite end is
\begin{equation}
v_d = -\omega (L - d_p) = -\alpha \left(\frac{L - d_p}{d_p}\right) v_w.
\end{equation}
The negative sign denotes the reversed direction of motion (upwind). If $(L - d_p)/d_p > 1/\alpha$, the magnitude $|v_d|$ exceeds $v_w$. As in the gearbox, the ratio depends only on geometry and not on absolute dimensions.

Once again, the pivot must be fixed to an external support so that it can supply the necessary reaction forces and torque. Without this constraint the lever would simply be carried along with the wind, producing little or no rotation and therefore no kinematic amplification---just as a vehicle lacking a no-slip ground contact cannot sustain motion faster than the surrounding air.

The lever analogy restates these same principles geometrically: the lever-arm
ratio sets the speed amplification, the pivot location sets the direction, and
the fixed support supplies the necessary reaction forces and torque.

\subsection{The Popescu Glide: No Propulsion from Internal Forces Alone}

The preceding analogies show that sustained propulsion requires an external
reaction through which momentum can be exchanged with a second medium. Without
such a reaction, internal motions merely redistribute momentum within the
system. To make this explicit, consider the following simple thought
experiment~\cite{popescu_talk_2024}.

Imagine Alice standing on a dock while Bob sits in a small boat that is free to
drift horizontally with negligible resistance (see Fig.~\ref{fig:popescu-glide}). Bob attempts to reach Alice by
walking from the stern to the bow. As he walks forward, the boat slides
backward. Because no external \emph{horizontal} forces act on the combined
Bob--boat system, its center of mass remains fixed, even though Bob expends internal energy while walking: Bob's forward motion is
exactly offset by the boat's recoil.

Crucially, unlike in the wind-cart configuration, the Bob--boat system in this
idealization lacks any \emph{traction} (no-slip) constraint to an external
body/medium. It can ``glide'' because there is no fixed reaction surface that
can supply an external horizontal impulse. In the wind-cart problem, by
contrast, the wheel--ground interface enforces a rolling-without-slip
constraint (instantaneous contact-point velocity zero in the ground frame),
which permits the drivetrain to exchange momentum with the ground and thereby
achieve sustained propulsion relative to the environment. In this sense the
Popescu Glide resembles a free-flying aircraft, which interacts essentially
with only a single medium, as discussed in Sec.~\ref{wind:whynotintheair}.

Let Bob have mass $m$ and the boat mass $M$. If Bob walks a distance $l$ relative to the boat, the boat must move backward by a distance $\Delta x_\text{boat}$ such that the system's center of mass does not shift:
\begin{equation}
m\,\Delta x_\text{Bob} + M\,\Delta x_\text{boat} = 0.
\end{equation}
Since Bob's displacement relative to the water is the sum of his motion relative to the boat and the boat's recoil,
\begin{equation}
\Delta x_\text{Bob} = l + \Delta x_\text{boat},
\end{equation}
we obtain
\begin{equation}
m(l + \Delta x_\text{boat}) + M\,\Delta x_\text{boat} = 0.
\end{equation}
Solving for the boat's recoil gives
\begin{equation}
\Delta x_\text{boat} = -\frac{m}{m+M}\,l.
\end{equation}
The negative sign indicates the boat moves backward. Bob's resulting progress toward Alice, measured relative to the water, is therefore
\begin{equation}
\Delta x_\text{Bob} = l - \frac{m}{m+M}\,l = \frac{M}{m+M}\,l.
\end{equation}
Thus, Bob advances only a fraction of the distance he walks, set by the mass ratio \(M/(m+M)\). Only in the limit \(M\to\infty\)---an effectively immovable boat, i.e., one rigidly constrained to an external support---does he make the full progress \(\Delta x_\text{Bob}\to l\).

Thus, without an external horizontal impulse, internal motions merely
redistribute momentum within the combined system and cannot produce net
translation relative to the environment.

The same logic applies to a wind-driven vehicle: Its drivetrain and propeller
can only produce sustained propulsion \emph{if the vehicle can exert forces on
a second medium}---ground or water---that supplies the required external
reaction forces (for example through traction at a wheel--ground contact or
through hydrodynamic sideforce). Without this external coupling, the vehicle's
internal motions would largely cancel through recoil of the rest of the system,
and no sustained translation relative to the environment would result.

This conclusion is independent of where the energy originates: it holds both
for \emph{external} energy inputs (e.g., aerodynamic work done by the wind on a
rotor) and for \emph{internal} energy sources (e.g., Bob's chemical energy
expended while walking). Energy expenditure alone is not sufficient for net
propulsion of a closed system; an external momentum exchange (an external
impulse) is required.

\begin{figure}[htbp]
\centering
\begin{tikzpicture}[
    scale=0.7, transform shape,
    line cap=round,
    line join=round,
    >=Stealth,
    font=\normalsize
  ]
  \useasboundingbox (-5.2,-0.8) rectangle (6.6,9.4);

  \begin{scope}[yshift=5.2cm]

    \node[above] at (0,-1) {(a) Initial configuration};

    \draw[thick,decorate,decoration={snake,amplitude=1.5pt,segment length=20pt}]
      (-5.0,0) -- (6.0,0);

    \begin{scope}[shift={(-3.0,0.4)}]
      \draw[thick, fill=white] (-1.8,1.0) -- (-1.5,0) -- (1.2,0) -- (1.5,1.0) -- cycle;

      \draw[thick] (-0.15,2.4) circle (0.25);     
      \draw[thick] (-0.15,2.15) -- (-0.15,1.0);   
      \draw[thick] (-0.15,1.8) -- (-0.6,1.4);     
      \draw[thick] (-0.15,1.8) -- (0.3,1.4);      
      \draw[thick] (-0.15,1.0) -- (-0.6,0.3);     
      \draw[thick] (-0.15,1.0) -- (0.3,0.3);      

      \node[above] at (-0.15,2.8) {Alice};
    \end{scope}

    \begin{scope}[shift={(1.55,0.4)}]
      \draw[thick, fill=white] (-1.3,1.0) -- (-1.0,0) -- (1.7,0) -- (2.0,1.0) -- cycle;

      \draw[thick] (0.95,2.4) circle (0.25);      
      \draw[thick] (0.95,2.15) -- (0.95,1.0);     
      \draw[thick] (0.95,1.8) -- (0.50,1.4);      
      \draw[thick] (0.95,1.8) -- (1.40,1.4);      
      \draw[thick] (0.95,1.0) -- (0.50,0.3);      
      \draw[thick] (0.95,1.0) -- (1.40,0.3);      

      \node[above] at (0.95,2.8) {Bob};
    \end{scope}

  \end{scope}

  \begin{scope}[yshift=0cm]

   \node[above] at (0,-1) {(b) Final configuration};

    \draw[thick,decorate,decoration={snake,amplitude=1.5pt,segment length=20pt}]
      (-5.0,0) -- (6.0,0);

    \begin{scope}[shift={(-3.0,0.4)}]
      \draw[thick, fill=white] (-1.8,1.0) -- (-1.5,0) -- (1.2,0) -- (1.5,1.0) -- cycle;

      \draw[thick] (-0.15,2.4) circle (0.25);     
      \draw[thick] (-0.15,2.15) -- (-0.15,1.0);   
      \draw[thick] (-0.15,1.8) -- (-0.6,1.4);     
      \draw[thick] (-0.15,1.8) -- (0.3,1.4);      
      \draw[thick] (-0.15,1.0) -- (-0.6,0.3);     
      \draw[thick] (-0.15,1.0) -- (0.3,0.3);      

      \node[above] at (-0.15,2.8) {Alice};
    \end{scope}

    \begin{scope}[shift={(2.45,0.4)}]
      \draw[dashed, thick] (-1.3,1.0) -- (-1.0,0) -- (1.7,0) -- (2.0,1.0) -- cycle;

      \draw[dashed, thick] (-0.35,2.4) circle (0.25);     
      \draw[dashed, thick] (-0.35,2.15) -- (-0.35,1.0);   
      \draw[dashed, thick] (-0.35,1.8) -- (-0.80,1.4);    
      \draw[dashed, thick] (-0.35,1.8) -- (0.10,1.4);     
      \draw[dashed, thick] (-0.35,1.0) -- (-0.80,0.3);    
      \draw[dashed, thick] (-0.35,1.0) -- (0.10,0.3);     

      \node[above] at (-0.35,2.8) {Bob};
    \end{scope}

    \draw[->,thick] (1.5,2) -- (0.9,2);
    \draw[->,thick] (1.5,2) -- (0.6,2);
    \node[above] at (1,2.2) {$\Delta x_{\mathrm{Bob}}$};

    \draw[->,thick] (4.9,0.8) -- (5.55,0.8);
    \draw[->,thick] (4.8,0.8) -- (6.05,0.8);
    \node[above] at (5.4,1) {$\Delta x_{\mathrm{boat}}$};

  \end{scope}

\end{tikzpicture}
\caption{Schematic of the ``Popescu Glide'' thought experiment.
(a) Initial configuration.
(b) Final configuration after Bob attempts to move toward Alice: Bob moves
leftward within his frictionless boat while the boat recoils rightward.
Without an external interaction, such as paddling in the water, the center of
mass of the Bob--boat system remains fixed, and he cannot achieve net propulsion.}
\label{fig:popescu-glide}
\end{figure}

Crucially, the Popescu Glide illustrates that without an external reaction
surface---an \emph{Archimedean point} in the sense of the dictum
``Give me a place to stand on, and I will move the earth''~\cite[p.~15]{dijksterhuis_archimedes_2014}---internal motions alone cannot change the
center-of-mass motion of the system relative to the environment. In the absence
of such an external constraint, internal actuation merely redistributes
momentum within the system.

Just as the gearbox and lever analogies require a fixed support that can supply
reaction forces and torque, a wind-driven vehicle requires contact with a
second medium (ground or water) that can provide the needed external reaction:
on land this is supplied by traction enforcing rolling without slip, while on
water it is mediated by hydrodynamic forces and wake formation. In this sense,
the external contact is not an accessory but an essential element that enables
sustained momentum exchange between air and ground (or water) and thereby makes
faster-than-the-wind travel possible.

Kirk Thomas McDonald has pointed out that, under more realistic assumptions,
Bob may be able to reduce the separation to Alice by exploiting effects that are
absent in the idealized ``glide'' model~\cite{McDonald2026priv}.

One possibility is for Bob to jump off the boat. If the dock is close enough,
Bob can jump from the bow and land on the dock. During the push-off, the
combined Bob--boat system still conserves horizontal momentum (so the boat
recoils), but once Bob makes contact with the dock the situation is no longer a
closed system: the dock provides an external impulse that allows Bob to come to
rest on shore.

Another issue is hydrodynamic drag (water ``friction''). In reality the boat
experiences drag from the water. When Bob walks forward relative to the boat,
the boat recoils backward, but the water exerts a drag force opposite this
recoil (i.e., a forward force on the boat). This external horizontal force
prevents the boat from recoiling far enough to keep the Bob--boat center of
mass fixed in the water frame. In this process momentum is transferred to the
surrounding water, and mechanical energy is dissipated in viscous losses and
wake formation (in the absence of currents, no net energy is extracted from the
water).

\section{Pedagogical Use: Teaching Levels, Learning Goals, Misconceptions, and Activities}
\label{sec:pedagogy}

The phenomenon and its analogies lend themselves to rich instruction from secondary school through undergraduate mechanics and laboratory courses. This section provides a concise, ready-to-use guide.

\subsection{Target levels and prerequisites}

\begin{itemize}
\item Secondary/High school (ages ~15--18): Newton's laws, frames of reference, work--energy, static vs kinetic friction, basic gear ratios or levers.
\item Introductory undergraduate physics/engineering: Rates of mechanical energy transfer, efficiency, drag scaling, turbines vs fans, idealizations vs nonidealities.
\item Upper-level mechanics/fluids: Constraint forces and power flow, momentum exchange between media, efficiency and performance limits, modeling/simulation.
\end{itemize}

\subsection{Learning goals}

By the end of instruction, students should be able to:
\begin{enumerate}
\item Distinguish airspeed, ground speed, and apparent wind; compute frame-dependent velocities.
\item Explain how static friction can transmit force without doing work in the ground frame; identify where mechanical energy enters and leaves the system and the corresponding rates of transfer.
\item Describe the vehicle as a mechanical transformer between two media; draw qualitative rate-of-energy-transfer (power-flow) diagrams for downwind and upwind operation.
\item Use the gearbox/lever analogies to predict velocity amplification and force reduction; derive and interpret the velocity-amplification factor $v/v_w = r_3/(r_3-r_2)$ in Eq.~\eqref{eq:amp-v-vw-ratio}.
\item Explain why faster-than-the-wind travel requires two media in relative motion and an external reaction surface (e.g., traction enforcing a no-slip ground contact) that enables momentum exchange between them.
\item Predict qualitative effects of changing gear ratio or blade pitch on achievable speed ratio and thrust.
\item Diagnose and correct common misconceptions (Section~\ref{sec:misconceptions}).
\end{enumerate}

\subsection{Typical conceptions and misconceptions}
\label{sec:misconceptions}

\begin{itemize}
\item ``If the cart matches the wind speed, the apparent wind is zero, so it can't accelerate further.''
\\Resolution: At $v=v_w$ the propeller is still actively driven by the wheels (fan mode downwind). Power comes from the relative motion of air and ground transmitted through the drivetrain; the prop adds energy to the air while the cart gains speed.

\item ``The ground must do positive work on the cart to speed it up.''
\\Resolution: With rolling without slip, the instantaneous wheel--ground contact point has zero velocity in the ground frame, so the ground does no work in the ideal limit. It nevertheless supplies tangential reaction forces and torque (traction) that allow the drivetrain to transfer momentum between air and ground; the net power ultimately comes from the wind (air motion relative to the ground).

\item ``This violates energy conservation.''
\\Resolution: No violation occurs. The mechanism couples two media and is not a single-turbine extraction from a uniform airstream.

\item ``A plane can't do this, so a cart can't either.''
\\Resolution: A plane operates in a single medium. The cart leverages two media with a mechanical coupling through the ground.

\item Frame confusion: mixing air-frame and ground-frame velocities and rates of energy transfer.
\\Resolution: Require explicit statements of the chosen frame and draw power-flow arrows with signs and reference frames.
\end{itemize}

\subsection{Classroom activities and exercises}

\begin{itemize}
\item Lever analogy (Fig.~\ref{fig:lever-downupwind}): Clamp a meter stick to a bench to form a fixed pivot. Apply a gentle push at distance $d_s$ from the pivot and compare the tip speed with the speed at the push point; then move the pivot between the push point and the tip to reverse the direction of motion (upwind analogue). Learning focus: kinematic amplification and the role of an external constraint (reaction forces and torque at the pivot).
\item Popescu Glide: A student standing on a low-friction cart or skateboard walks forward; the cart rolls backward. Discuss center-of-mass motion and why, in the absence of an external reaction surface, internal motions cannot produce net translation relative to the room. Variation: have the student pull briefly on a wall-mounted rope to illustrate how an external constraint (an external impulse) changes the outcome.
\end{itemize}

\subsection{Conceptual and quantitative exercises}

\begin{enumerate}
\item Conceptual (frames): In the ground frame, explain how the ground does no work on the cart in the ideal rolling limit, yet the cart accelerates. Where does the power come from?

\item Power balance at the ``crossover'': Discuss why, even when $v=v_w$ downwind, the propeller can still produce thrust. Identify the element acting as the power sink/source at that instant.

\item Upwind variant: Using the top--bottom inverted negative-transmission gearbox, explain why the cart can move upwind. How does the sign of $v/v_w$ change, and why is the upwind magnitude not generally the same as the downwind magnitude?
\end{enumerate}

\subsection{Assessment ideas and rubrics (brief)}

\begin{itemize}
\item Exit ticket: ``In one or two sentences, explain why the ground can provide a force yet do no work on the cart.''
\item Short quiz: Multiple-choice on frames and signs of power; one free-response drawing a rate-of-energy-transfer (power-flow) diagram.
\item Lab report: Methods (measurement of $v_w$ and $v$), uncertainty estimates, and a discussion relating data to the gearbox/lever analogies.
\end{itemize}

In summary, these activities concretize otherwise abstract ideas---constraints
and reaction forces, power flow, and reference-frame reasoning---and make the
mechanism underlying the counterintuitive result accessible in the classroom
with modest equipment.

\section{Conclusion}

The seemingly paradoxical ability of a vehicle to travel directly downwind
faster than the wind is, at its core, a lucid demonstration of classical
mechanics rather than an exception to it. The apparent mystery dissolves once
we shift perspective: the vehicle is not merely being \emph{pushed} by the wind
but functions as a coupled machine that extracts energy from the \emph{relative
motion} of two media, namely air and ground (or water).

The mechanical analogies developed here clarify this interplay from
complementary viewpoints. The \emph{gearbox} illustrates how relative motion can
be transformed and amplified through fixed ratios, while the \emph{lever}
expresses the same principle geometrically through leverage. Both analogies
highlight the essential requirement of an external constraint---provided in
practice by traction at a ground contact (or, less rigidly, by hydrodynamic
reaction on water)---which supplies reaction forces and torque and thereby
permits sustained momentum exchange between the two media. The \emph{Popescu
Glide} thought experiment reduces this point to its purest form: without an
external reaction surface, internal motions merely redistribute momentum within
the system and cannot produce net translation relative to the environment.


The role of a traction constraint in faster-than-the-wind travel is closely
related to the classical Archimedean theme. In the traditional accounts
\cite[pp.~14--17]{dijksterhuis_archimedes_2014}, Archimedes demonstrates that a
sufficiently clever machine---a lever, a tackle (polyspaston/trispaston), or a
windlass driven through gears and even an endless screw---can ``magnify''
force and thereby move an enormous load (such as a ship) with apparently small
effort. The essential caveat, already implicit in the dictum
\emph{``Give me a place to stand on, and I will move the earth''} mentioned earlier
\cite[p.~15]{dijksterhuis_archimedes_2014}, is that mechanical advantage cannot
be exploited to move a system relative to its environment without a fixed
external support that supplies the reaction forces and torque: the machine must
push \emph{against something} that does not simply recoil with it.

In the present context, the wind-cart drivetrain plays the role of Archimedes'
``leveraging'' device: it is a mechanical transformer that trades force for
speed (or vice versa) while ideally conserving power. However, the drivetrain
alone cannot produce sustained motion relative to the environment. The required
Archimedean point is supplied by the external constraint at the ground contact:
in the ideal rolling-without-slip limit the instantaneous contact-point
velocity vanishes in the ground frame, so the ground can supply a horizontal
reaction (static-friction traction) and a balancing torque while doing no work
at the contact. This is precisely what the Popescu Glide lacks: without a fixed
reaction surface, internal actuation merely redistributes momentum within the
system, leaving its center-of-mass motion unchanged. (As emphasized in the
historical accounts, idealized demonstrations neglect frictional losses; in
practice aerodynamic drag, rolling resistance, and incipient slip bound the
attainable speed ratios.)

The gearbox analogy makes this identification explicit. The upper rack
represents the moving air, the lower rack represents the fixed ground, and the
gear train is the ``machine.'' The velocity-amplification factor
$v/v_w = r_3/(r_3-r_2)$ in~\eqref{eq:amp-v-vw-ratio} is the analogue of a lever-arm ratio or a tackle's
reeving ratio: it achieves a larger output speed at the cost of a reduced
transmitted force. In Archimedes' canonical lever setting the emphasis is
often the dual statement: a reduced effort force is obtained at the cost of a
larger displacement on the effort side; which side exhibits the ``greater
motion'' depends on which end is taken as input. This kinematic amplification
becomes physically operative only because one interface (the ground rack)
provides the non-recoiling support---the Archimedean point---that closes the
force/torque balance of the drivetrain and permits sustained momentum exchange
between air and ground.

For physics educators, these analogies provide an accessible path to teaching a
counterintuitive yet conceptually rich phenomenon. They reveal that
faster-than-the-wind motion is not paradoxical but inevitable once one
appreciates how two interacting media can be coupled to generate geared-up
motion. The broader lesson is that in any physical system, when two
environments are in relative motion, a suitably designed mechanism can use one
to propel itself through the other---transforming a perceived impossibility
into an elegant example of mechanical ingenuity.

\begin{acknowledgments}
The author thanks Sandu Popescu for a helpful communication regarding his inspiring boat analogy, and Kirk Thomas McDonald for stimulating and constructive criticism of earlier versions of the manuscript. Any remaining errors or misconceptions are the author's responsibility.
This research was funded in whole or in part by the \textit{Austrian Science Fund (FWF)} [Grant \href{https://doi.org/10.55776/PIN5424624}{DOI:10.55776/PIN5424624}].
The author acknowledges TU Wien Bibliothek for financial support through its Open Access Funding Programme.

This text was partially created and revised with assistance from large language models. All content, ideas, and prompts were provided by the author.
\end{acknowledgments}

\bibliography{svozil}

\end{document}